\documentclass[12pt,preprint]{aastex}
\begin{document}

\title{GRB 070714B - Discovery of the Highest Spectroscopically Confirmed Short Burst Redshift}
\author{J. F. Graham\affil{Space Telescope Science Institute \\ Johns Hopkins University}}
\author{A. S. Fruchter\affil{Space Telescope Science Institute}}
\author{A. J. Levan\affil{University of Warwick}}
\author{A. Melandri\affil{Liverpool John Moores University}}
\author{L. J. Kewley\affil{University of Hawaii}}
\author{E. M. Levesque\affil{University of Hawaii}}
\author{M. Nysewander\affil{Space Telescope Science Institute}}
\author{N. R. Tanvir\affil{University of Leicester}}
\author{T. Dahlen\affil{Space Telescope Science Institute}}
\author{D. Bersier\affil{Liverpool John Moores University}}
\author{K. Wiersema\affil{University of Leicester}}
\author{D. G. Bonfield\affil{NASA Goddard Space Flight Center}}
\author{A. Martinez-Sansigre\affil{Max Planck Institute for Astronomy}}

\author{\today}

\begin{abstract}

We detect the optical afterglow and host galaxy of GRB 070714B.  Our observations of the afterglow show an initial plateau in the lightcurve for approximately the first 5 to 25 minutes, then steepening to a powerlaw decay with index $\alpha = 0.86 \pm 0.10$ for the period between 1 to 24 hours post burst.  This is consistent with the X-ray light-curve which shows an initial plateau followed by a similar subsequent decay.  At late time, we detect a host galaxy at the location of the optical transient.  Gemini Nod \& Shuffle spectroscopic observations of the host show a single emission line at 7167 {\AA} which, based on a \textit{grizJHK} photometric redshift, we conclude is the 3727 {\AA} [O II] line.  We therefore find a redshift of $z=0.923$.  This redshift, as well as a subsequent probable spectroscopic redshift determination of GRB 070429B at $z=0.904$ by two other groups, significantly exceeds the previous highest spectroscopically confirmed short burst redshift of $z=0.546$ for GRB 051221.   This dramatically moves back the time at which we know short bursts were being formed, and suggests that the present evidence for an old progenitor population may be observationally biased.

\end{abstract}
\keywords{gamma rays: bursts}

\section{Introduction}

Although evidence of two classes of gamma-ray bursts has existed for over twenty-five years \cite{Mazets, Norris} it was only with the work of \cite{Kouveliotou} 15 years ago that gamma ray bursts  were widely recognized as being divided into two types based on duration and spectral slope: Short and hard GRBs (SGRBs) and Long and soft GRBs (LGRBs).  \cite{Katz} then used the different $\langle V/V_{max} \rangle$ \citep{Schmidt} values between the SGRB and LGRB populations to claim a different distance distribution between the SGRB and LGRB populations,  and to suggest that their formation mechanisms and thus their progenitors are likely different.  They proposed that SGRBs were the product of compact object mergers (a possibility already noted earlier by, e.g. \citealt{Blinnikov, Paczynski, Goodman, Eichler}).  More recently the discovery of short burst afterglows (c.f. \citealt{HjorthNature, FoxNature, BergerNature, Soderberg2006}) has shown significant differences between the host galaxy populations of the two types.

Long bursts are predominantly associated with a particular galaxy type, faint blue irregular galaxies \citep{Fruchter1999, Fruchter, LeFlochblue, LeFlochblue2002}, and low metallicity environments in general \citep{Fruchter, Stanek2007, Modjaz2008}.  Additionally, they show a strong preference for occurring not only in starforming galaxies \citep{Fruchter1999, Christensen, LeFloch2006} but in the brightest, and hence likely most star forming, regions of their hosts \citep{Fruchter}.  This combined with their being frequently associated with subsequent type Ic supernova events \citep{Stanek, Hjorth2003, Woosley} has provided a coherent picture of the LGRB progenitor system.

With short bursts a similar understanding is proving more elusive.  While long bursts possess afterglows that are the most intense and among the most luminous astronomical phenomena \citep{Kann, Bloominpress}, short bursts tend to have fainter afterglows \citep{Kanninpress}.  Thus, absorption spectroscopy (which has provided a critical insight into the immediate progenitor environment of long bursts, e.g. \citealt{Prochaska, Vreeswijk}) of a short burst afterglow has, to date, not been successfully obtained (despite attempts \citealt{Stratta, Piranomonte}).  Until an SGRB afterglow absorption spectrum is obtained, almost all our knowledge of the burst environment, particularly the ISM, must be derived from observations of the host galaxies associated with the bursts \citep{Bergerinpress}.

Short burst host galaxies show a much greater diversity than those of long bursts.  While the initial SGRB host associations were in elliptical galaxies (SGRB 050509B, \citealt{050509b}; SGRB 050724, \citealt{BergerNature}) leading to claims that an old progenitor population was required, subsequent observations have suggested that this was due to small number statistics.  The occurrence of short bursts is now associated with all types of galaxies from elliptical to star-forming dwarf and is not obviously dependent on the host galaxy's (current) star formation \citep{Gehrels, BergerNature}.  This is most likely indicative of a delay between progenitor system formation and burst occurrence \citep{Nakar} and along with the absence of any correlating supernova emission \citep{Hjorth, Kanninpress} makes single massive star core collapse an unlikely formation method.

Due to the greater diversity of short burst host galaxies types, and the greater complexity in deriving a useful analysis of galaxy properties for some of these types, considerably greater complications are encountered in short burst host galaxy analysis requiring careful consideration of observational biases.   Also, while the delay between the formation of short burst progenitors and the occurrence of the burst might allow binary progenitors to leave their host galaxies (c.f. \citealt{Belczynski}), this should not used as a panacea for explaining away host burst pairings with unacceptably large separations, especially considering that many SGRBs do not have bright hosts \citep{Berger2007}.

Compact object mergers, e.g. neutron star-neutron star or neutron star-black hole \citep{Eichler, Davies, Lee}, are presently considered the preferred short burst progenitor candidate (c.f. \citealt{Gehrels, 050509b}).  Direct conclusive evidence will most likely come from gravitational wave detections of SGRB merger events perhaps from the next generation of gravitational wave detectors.  Frequency estimates for a LIGO II and \emph{Swift} concurrent detection are in the range of one approximately every 3 to 10 years (\citealt{Seto} - for a neutron star-black hole and neutron star-neutron star merger respectively).  Until then competing possibilities that satisfy the aforementioned criteria, such as millisecond pulsars with extremely strong magnetic fields \citep{Usov}, collapse of neutron stars into black holes in binary systems \citep{Dermer}, and magnetar production via white dwarf-white dwarf mergers \citep{Levan}, cannot be ruled out.  However host observations may provide the best opportunity to constrain various formation models.

Here we report on photometry and spectroscopy of the host galaxy of GRB 070714B, which lead us to conclude that the host is a moderately star-forming galaxy at a redshift of z=0.923 and suggesting that the present evidence for an old progenitor population may be, at least partially, observationally biased.  The spectroscopy and redshift discussed here was originally reported in GCN 6836 \citep{GCN6836}.

\section{Observations and Data Reduction}

\subsection{\emph{Swift}}

GRB 070714B was initially detected by the Burst Alert Telescope (BAT) on the 14th of July 2007 at 4:59:29 UT \citep{GCN6620} on the NASA \emph{Swift} spacecraft \citep{swift}.  The gamma ray emission consisted of several short spikes with a collective duration of 3 seconds followed approximately twenty seconds later by fifty seconds of softer emission.  The main component also shows a small spectral lag \citep{GCN6529}.  This emission is similar to previous short bursts including GRB 050724 \citep{GCN6623}, and places this burst securely in the short category.

Rapid localization with the BAT instrument allowed the \emph{Swift} satellite to slew its additional instruments onto the source.  The onboard X-Ray Telescope (XRT) detected a new fading X-ray source 35 arc seconds away from the final ground processed BAT position \citep{GCN6627}.  While the onboard UltraViolet Optical Telescope (UVOT) did not initially detect any new source within the XRT error circle, down to an unfiltered limit of 20.4 mag \citep{GCN6632}, a late time reanalysis via coadding all prompt data detected a 4.5 sigma optical transient with an unfiltered magnitude of 20.95 $\pm$ 0.23 \citep{GCN6689}.

\subsection{Afterglow follow-up}

Our first observations of GRB 070714B were undertaken with the Liverpool Telescope starting roughly 12 minutes after the burst. A series of short 10s $r$-band exposures followed by longer 120s integrations in  \textit{riz} were obtained.  We discovered an optical afterglow at RA: 03$^{h}$51$^{m}$22.2$^{s}$ Dec: +28$^{\circ}$17$^{m}$51.4$^{s}$ (J2000) with 0.5" error \citep{GCN6621} which places it within the XRT error circle \citep{GCN6627}.  The afterglow was only marginally detected in the short exposures and these were stacked to improve accuracy of the photometry.  Additional observations were obtained the following night at the William Herschel Telescope in the R-band \citep{GCN6630}.  Subsequent observations with Gemini North detected no afterglow emission (see section \ref{opticalmags}).

Photometry of the afterglow was taken using apertures with a radius equal to the FWHM of a stellar point source in each of the images. Our WHT R-band observations were photometred into the \textit{r} band by utilizing the photometric transformations of \cite{Jester}; however, given the modest signal to noise in the detection, this transformation is not the dominant source of uncertainty.  Absolute photometric calibration was obtained by using field stars to scale the zero point values to the Gemini field calibration exposures as described in Section \ref{opticalmags}.  A log of the early photometry of the afterglow (and host contribution) is show in Table \ref{phot-ot}.
\begin{table}
\begin{center}
\begin{tabular}{|c|c|c|c|c|c|c|c|}
\hline
Date (UT) & $\Delta$T (days) & Telescope & Band  & Exposure Time &  mag  $\pm$ error\\   
\hline
 14.21685     & 0.0088          &    Liverpool    &        \textit{r} &     6$\times$10  &  20.30 $\pm$   0.18 \\
 14.21917     & 0.0112          &    Liverpool    &        \textit{r} &     120   &  20.14 $\pm$   0.11\\
 14.22030     & 0.0123          &    Liverpool    &        \textit{i} &     120   &  19.88 $\pm$   0.17 \\
 14.22473     & 0.0168          &    Liverpool    &        \textit{r} &     120   &  20.28 $\pm$   0.25 \\   
 14.22625     & 0.0183          &    Liverpool    &        \textit{r} &     120   &  20.27 $\pm$   0.23 \\
 15.19236     & 0.9844          &    WHT    &         \textit{R} &      8$\times$300 &  24.11 $\pm$  0.21\\
\hline
\end{tabular}
\caption{Observations of the GRB 070714B afterglow. (The \textit{r} is SDSS magnitude whereas \textit{R} is Vega magnitude).  The host contribution has not been removed. \label{phot-ot}}
\end{center}
\end{table}
\subsection{Host Galaxy Optical Photometry} \label{opticalmags}

Optical imaging of the host galaxy was obtained in \textit{g}, \textit{r}, \textit{i}, and \textit{z} bands with the GMOS instrument on Gemini North between the nights of July 16$^{th}$ and 26$^{th}$, 2007.  The number of individual exposures and nights of observation are given in Table \ref{mags}.  An individual exposure time of 300 seconds was used for all bands.  With the exception of the \textit{z} band observations, this imaging was collected in non-photometric conditions.  Due to the narrow window between the object rising and dawn, optical observations were performed shortly before and often extending slightly past astronomical twilight, at airmasses just below and sometimes slightly above 2, and had to be spread across several nights.

\begin{figure}[h]
\begin{center}
\includegraphics[width=1\textwidth]{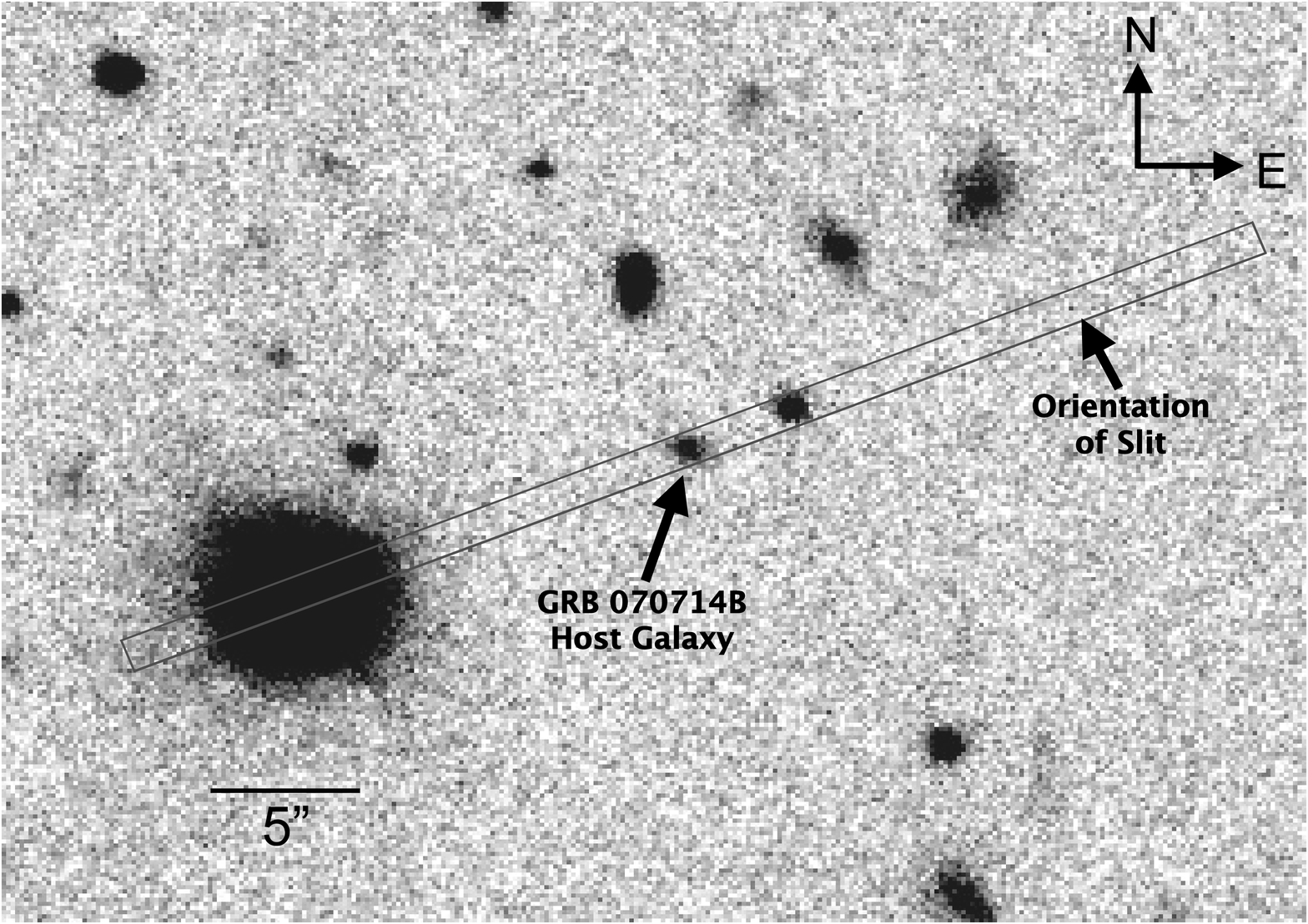}
\caption{Combined \textit{i} band image of the GRB 070714B host and surrounding field.  The host and the slit orientation used in the spectroscopy are annotated.  \label{slit}}
\end{center}
\end{figure}

Photometric calibration observations were subsequently obtained of the object field and a standard field with GMOS on Gemini South on the night of September 25$^{th}$.  Object field observations were performed with two 60-second exposures in all bands.  Two standard exposures were taken in each band, encompassing 4 stars (41, 42, 112, and 115) in selected area 95 of \cite{Landolt1992}, of 8 seconds duration in \textit{g}, \textit{i}, and \textit{z} and 6 seconds in \textit{r}.  The final combined \textit{i} band image of the object field is shown in Fig.\thinspace\ref{slit}.

Gemini optical imaging was reduced with the standard Gemini GMOS IRAF\footnote{IRAF is distributed by the National Optical Astronomy Observatories, which are operated by the Association of Universities for Research in Astronomy, Inc., under cooperative agreement with the National Science Foundation.} packages with fringing corrections applied in the \textit{i} and \textit{z} bands.  The variable weather conditions and approaching dawn complicated the combination of the individual exposures into an optimal combined image.

Images in each band were scaled to a common intensity, aligned, and combined using a normalized weighting of the square of the inverse intensity scaling times the inverse sky median.  Due to the near twilight nature of the observations the latter weighting was necessary to optimally correct for the worsening noise (due to increasing sky brightness) and corresponding change in signal to noise ratio.  A small alteration was made to the  ``imcoadd" Gemini IRAF task to introduce the weight.

In order to calibrate this science imaging the calibration exposures were similarly reduced but not combined.  SDSS magnitudes of the standard stars were used to determine zero point values of stars in each of the two standard calibration exposures in each band. (The SDSS survey magnitudes for the standards were used to maintain consistency in magnitudes systems with Gemini).  The average, for each band, was then used as the photometric zero point for the calibration images of the object field.  Field stars were then used to determine the scaling between the calibration and science exposures in each band.  

In the optical, a color correction between the GMOS-N filters and the SDSS filters is computed, using the values from \cite{Barr} in \textit{g} and \textit{r} and \cite{Inger} in \textit{r}, \textit{i}, and \textit{z} (the \textit{r} corrections are identical in both sources), and is as expected negligible.  A galactic extinction estimate of E(B-V) = 0.141 magnitudes from \cite{Schlegel} is then applied.  The final magnitudes of the host galaxy in each band are shown in Table \ref{mags}.

\begin{table}[h]
\begin{center}
\begin{tabular}{|c|c|c|c|c|c|c|c|}

\hline
band    &  raw mag& color corrected & dust extinction & date & number of & total time\\
        &  $\pm$ error  & mag & corrected mag & (UT) & exposures & (sec) \\
\hline
\textit{g}    &   25.79 $\pm$ 0.34   & 25.81 $\pm$ 0.34 & 25.28  $\pm$ 0.35 & 17.6  &  5  & 1500\\
\textit{r}    &   24.90 $\pm$ 0.21   & 24.93 $\pm$ 0.21 & 24.54 $\pm$ 0.22  & 19.6  & 11\tablenotemark{A}  & 3300\\
\textit{i}    &   23.97 $\pm$ 0.12   & 23.96  $\pm$ 0.12 & 23.67 $\pm$ 0.13  &  22.6 &  8 & 2400\\
\textit{z}    &   24.01 $\pm$ 0.13   & 23.99  $\pm$ 0.13 & 23.78 $\pm$ 0.13 &  26.6 &  8  &  2400\\
\hline
\textit{J}    &   22.27 $\pm$ 0.12   &  & 22.14 $\pm$ 0.12 & 26.6 &  20 &  1200\\
\textit{H}    &   22.28 $\pm$ 0.20   &  & 22.20 $\pm$ 0.20 & 25.6  &  17 & 1020\\
\textit{K}    &   21.13 $\pm$ 0.13   &  & 21.08 $\pm$ 0.13 & 24.6  &  16 &  960\\
\hline

\end{tabular} 
\caption{The optical and IR photometric magnitudes of the host galaxy in each observed band.  Note, only the IR observations were performed in photometric conditions, the optical data was subsequently calibrated.  (\textit{g}, \textit{r}, \textit{i}, and \textit{z} are SDSS magnitudes whereas \textit{J}, \textit{H}, and \textit{K} are Vega magnitudes).  Errors include both the statical and systematic component.  \label{mags}\label{irmags}}
\end{center}
{$^A$Three of these exposures were taken along with the \textit{g} band images on the 17$^{th}$ under comparatively poor conditions and comprise in total only 5\% of the weight when combined into the final \textit{r} band image.  (An additional two \textit{r} band images were taken on the 17$^{th}$ but were of such poor quality as to not be usable).}
\end{table}

\subsection{IR photometry}

We obtained \textit{JHK} near-infrared imaging of the host galaxy of GRB 070714B with the Gemini South NIRI instrument on the nights of July 24$^{th}$, 25$^{th}$ and 26$^{th}$, 2007.  The details of the observations are given in Table \ref{irmags}.  Single exposures were taken totaling one minute each in coadds of 5 x 12s, 6 x 10s and 5 x 12s, in \textit{J}, \textit{H}, and \textit{K}, for a total exposure time of 20, 17, and 16 minutes respectively.  The images were taken under photometric conditions and low airmass. 

Gemini near-infrared imaging was reduced with the standard Gemini NIRI IRAF package.  Images were normalized by a flat field, dark and sky subtracted, aligned and combined using the standard NIRI reduction guidelines.  In order to best account for variations in the sky, the images used for sky subtraction were based upon a rolling combination of exposures centered upon each image.  The near-infrared images were calibrated to unsaturated field stars with magnitudes available in the 2MASS catalogue, and were photometered in IRAF using a two FWHM aperture.  The final magnitudes of the host galaxy in each band are shown in Table \ref{irmags}.

\subsection{Optical spectroscopy}

Initial spectroscopic observations were obtained with the GMOS instrument on Gemini North on the night of the July 25$^{th}$.  Due to the drop in detector sensitivity long-ward of 9250 {\AA}, a central wavelength of 7250 {\AA} was selected yielding a spectral range of 5250 to 9250 {\AA}.  Further spectroscopy was conducted on the night of September 13$^{th}$ with the central wavelength shifted out to 7750 {\AA} in an effort to extend the spectral coverage to include a line possibility discussed in Section \ref{3727->5007}.

The R400 grating offers a reasonable compromise between spectral resolution (1.37 {\AA} pixel$^{-1}$) and width of coverage (about 4000 {\AA}) and was used both nights.  A 50 {\AA} dither in wavelength was also added both nights to ensure continuous spectral coverage across chip gaps and allow for easy removal of other chip based effects.  Due to the abundance of skylines in the spectral range the Nod \& Shuffle method was used.  The first and second night of spectroscopy consisted of four and six 10-minute Nod \& Shuffle exposures respectively.

In Nod \& Shuffle observing only the central third of a typical long slit is opened and the telescope is repetitively offset (nodded) between two positions on the slit while the CCD charge wells are simultaneously moved (shuffled) between the illuminated central region and the non-illuminated upper and lower thirds of the CCD.  These non-illuminated areas act as a storage region, buffering the data until the shutter is momentarily closed, the telescope offset, and the charge wells are shifted back into the illuminated region of the CCD.  While the nod only requires that the object be moved off its location on the image and that sky land there instead, in practice the object is moved along the slit to another location on the image.  Thus while one location is observing the object the other is observing the sky and no time is spent solely gathering sky data.  The result is a pair of images on the same CCD readout both containing the object spectrum yet located in a different place with regard to the slit's field of view on the image.

Nod \& Shuffle results in each CCD exposure being composed of multiple short exposures of the object interlaced with similar short exposures of the sky, each respectively stacked into separate images, each with an exposure time of half the total open shutter time of the Nod \& Shuffle exposure.  This produces much more rapid temporal sampling of the sky than with separate successive exposures and a much lower noise than with a set of multiple separate images each with their own readout thus eliminating the traditional tradeoff between the two concerns.  While the time observed on the object remains the total open shutter time of the exposure, the drawback is in the closed shutter time lost to moving the telescope between the two offset positions.  For a more detailed description of the Nod \& Shuffle process see \cite{Cuillandre} and \cite{Glazebrook} and for its use on Gemini see \cite{Glazebrookgemini} and \cite{Abraham}.

The individual spectroscopic exposures were reduced using the standard IRAF  ``Gemini.GMOS"  Nod \& Shuffle packages and process.  This is essentially the same as a conventional spectroscopic reduction except the two shuffled images on each exposure are subtracted from each other after bias subtraction and before flat fielding.  This yields a single image for each exposure, each containing two spectra, one of which is inverted, separated by the nod distance.  A custom Nod \& Shuffle dark was also used.

To optimize cosmic ray rejection for the relatively smaller number of Nod \& Shuffle spectra involved in this GRB host observation, we employ a deviation from the typical Nod \& Shuffle reduction process.  Since each reduced Nod \& Shuffle exposure contains one positive and one inverted spectrum (as described above), an inverted copy of the images is created such that (along with the originals) each spectrum is now present in an image without being inverted.  This resulted in eight and twelve 300 second images for the 7250 and 7750 {\AA} central wavelengths respectively.   For each central wavelength an offset between the positive spectra is determined and the images are then combined using this offset in a single step, thus optimizing the cosmic ray rejection of the images being combined by doubling their number.  (This is opposed to the conventional Nod \& Shuffle reduction process were the reduced Nod \& Shuffle exposures are combined, with cosmic ray rejection, and the positive and inverted spectra on combined output image are handled subsequently).  The two central wavelengths were then combined, weighted by their total exposure time.

Spectral extraction was performed with IRAF task ``apall" using a 10 pixel wide aperture in the spatial direction.  The continuum was too weak to allow automatic tracing of the aperture center.  However a tracing of a bright star, also present in the slit, showed insignificant deviation in the spatial direction along the length of the spectra.  Thus a fixed center aperture was used. \label{apall}

\begin{figure}[h]
\begin{center}
\includegraphics[width=.7\textwidth]{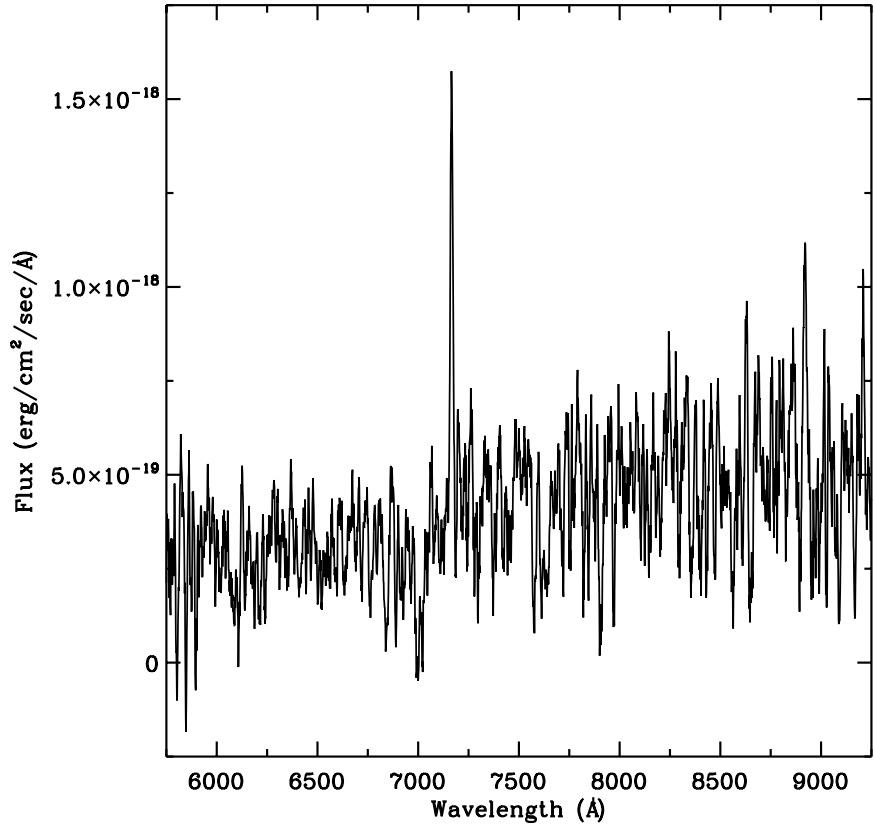}
\caption{Gemini \textit{GMOS} Nod \& Shuffle spectroscopy of the GRB 070714B host galaxy, smoothed with a 9 pixel boxcar function.  In order to maintain a consistent depth of coverage only the region of overlapping spectral coverage from the two central wavelengths is shown.  Only a continuum and a single emission line at 7167 {\AA} was detected. \label{spec}}
\end{center}
\end{figure}

\begin{figure}[h]
\begin{center}
\includegraphics[width=.7\textwidth]{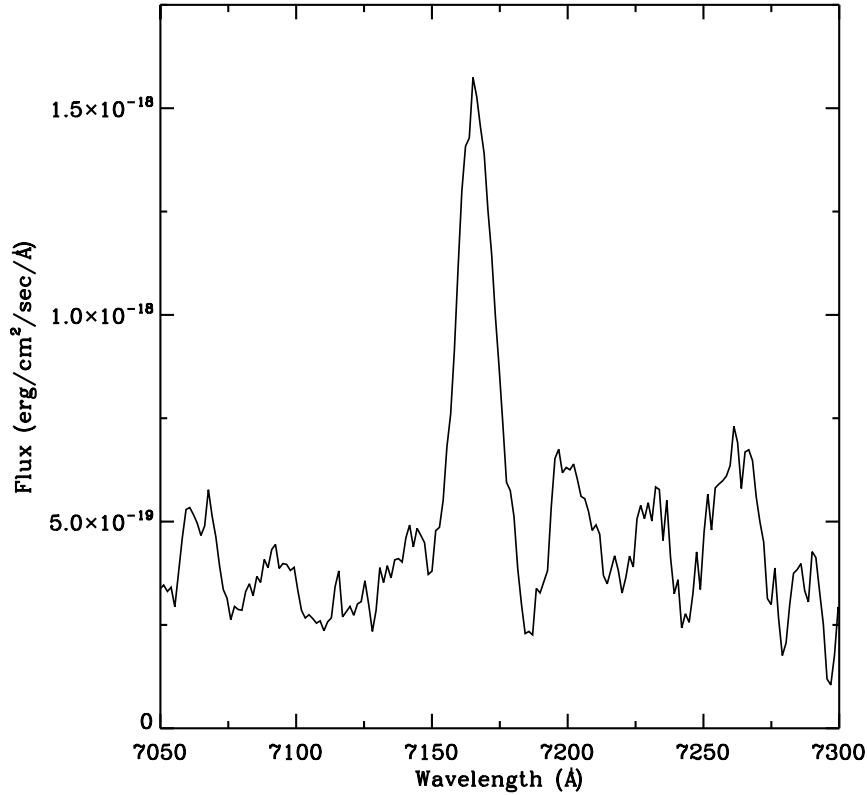}
\caption{An enlarged view of the single emission line detected at 7167 {\AA} and the sounding region from our Gemini \textit{GMOS} Nod \& Shuffle spectroscopy of the GRB 070714B host galaxy.  Smoothed, as in Fig.\thinspace\ref{spec}, with a 9 pixel boxcar function. \label{o2}}
\end{center}
\end{figure}

The process yielded a spectrum with a spectral resolution of 1.37 {\AA} per pixel and a spatial resolution of 0.15 arc seconds per pixel.  The spectrum is shown in Fig.\thinspace\ref{spec} and an enlargement of the region containing the spectral line in Fig.\thinspace\ref{o2}.  A single spectral line was observed at 7167 {\AA}.  Aside from a faint continuum no other spectral features were detected in the 5150 to 9900 {\AA} spectral range.

The spectrophotometric standard EG131 was observed immediately prior to the first epoch of spectroscopy and used to apply a relative flux calibration.  However, because the spectroscopy and standard were observed in non-photometric conditions, the relative flux calibrated spectrum was averaged (weighted by the product of the \textit{i} band filter transmissivity and the CCD spectral response) and then scaled such that this weighted average matches the calibrated \textit{i} band flux magnitude determined in Section \ref{opticalmags}.
\subsection{Near Infrared Spectroscopy}

Near infrared spectroscopy was performed on the host of GRB 070714B  with \textit{NIRSPEC} on the Keck II telescope on the night of October 23$^{th}$.  The object was too faint to be visible on the acquisition camera and was acquired by placing the slit on a bright star and using a position angle that aligned the slit onto the object.  Our observations consisted of eight 900 second exposures  in the NIRSPEC-3 filter, using a 0.76 arc-second slit, and giving a  spectral coverage from 1.15 to 1.33  $\mu$m and a spectral resolution  of 2.33 {\AA} per pixel.

Individual \textit{NIRSPEC} exposures were reduced using the standard  procedure described in the online documentation (from the NIRSPEC manual).  The  object's placement on the slit was dithered between two locations (four exposures in each), so that the combined image from each  placement could be subtracted from the other to remove sky features.   However this was complicated by the inadvertent placement of a field star on the slit at the sky subtraction position in one of the two  settings.   (These observations were unexpectedly obtained at the end  of a night when it became clear that time otherwise planned could be  used).  To compensate for this error the typical process of subtracting the images and (then aligning and) adding the spectra was altered.

First the sky effects were removed from individual exposures as best  as possible via fitting the sky in the spatial direction with a high order polynomial and then subtracting off the fit.  While this yielded a reasonably good removal of skylines  and generated an image that was useable via human inspection, it was still inferior to a traditional sky subtraction with a 2d sky image.  In particular a few faint features remain in the spectral direction, these have the appearance of a faint spectral continuum, but move with the slit, not the sky.  

As no sources were detected in the individual spectra,  the pixel offset needed to align the two dithered placements of the object on the CCD was determined from observations of another object using the  same setup and dither value earlier in the night.  The combined images for the two dithered placements were thus aligned and added.

Additionally another image was generated by reversing the direction of the offset when combining the images.  Since all defects in the images are shared between the two dither placements, this creates an image without any object data but with an identical set of these defects, however also containing the bright star.  Finally this image is subtracted from the one described in the previous paragraph yielding a subtracted image identical to that given in the traditional method.  However by blinking between this subtracted image and its operands one can visually exclude the remaining sky features.  No continuum or line for the object was detected.

To determine a rigorous upper limit value, the slit camera (SCAM) images were coadded for each dither position and then the coadded images were subtracted.  The resulting images was inverted, shifted (to align the subsequently inverted negative features with their positive counterparts) and then coadded with its original self to generate a combined image.  In addition to the bright star used for placing the list this uncovered two other stars on the final SCAM image.  This image was then astrometrically compared with the Gemini GMOS I band final image to determine the location of the target galaxy and the additional nearby galaxy on the SCAM image (see Fig.\thinspace\ref{slit}) and confirms that they were correctly placed on the slit.

A faint spectrum of the additional nearby galaxy described above was detected at the blue end of the spectrum.  This in addition to the spectrum of the bright star were used to determine the relation between the objects on the SCAM image and the 2d spectrum and interpolate the (expected) location of the object on the slit.  A blind extraction was then used to generate a 1d spectrum.  Again no continuum or line was apparent and based on the noise a three sigma upper limit on the possible line flux of $ 4.4 \times 10^{-17}$ erg s$^{-1}$ cm$^{-2}$ was determined.

\section{Analysis}

\subsection{Afterglow follow-up}

\begin{figure}[h]
\begin{center}
\includegraphics[width=1\textwidth]{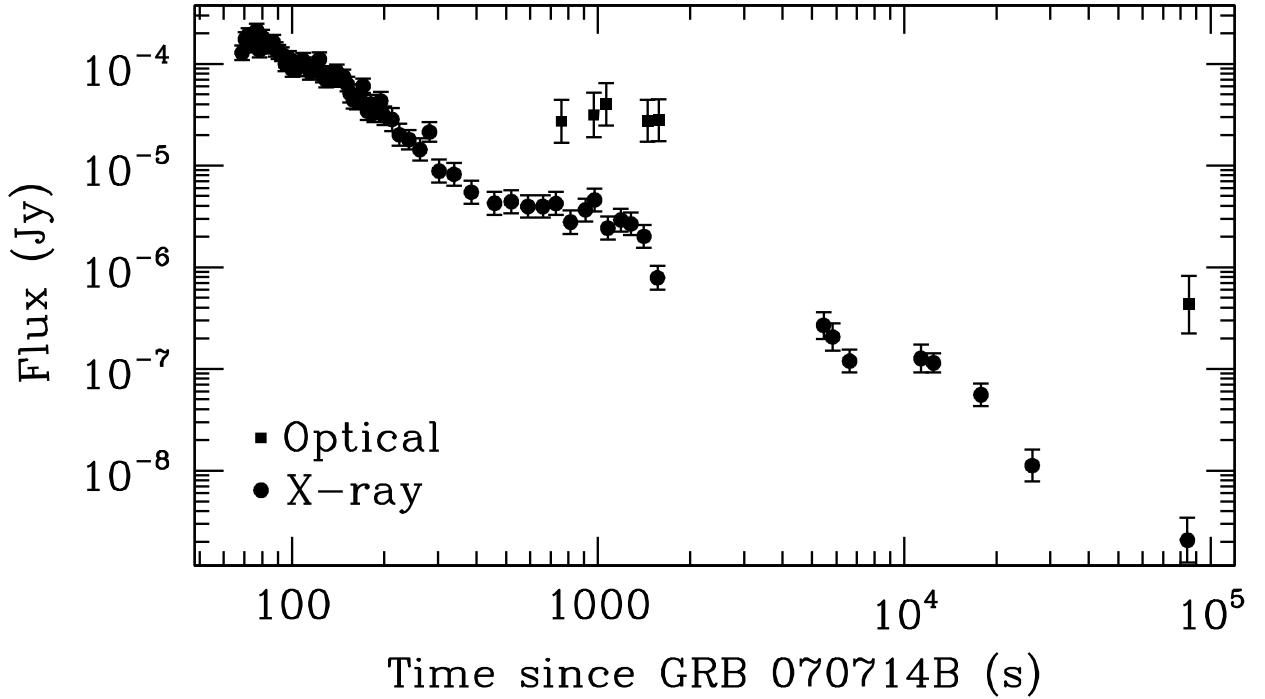}
\caption{Our optical observations of the GRB 070714B afterglow with the X-ray data over plotted.  The host contribution has been removed.  Note the coincident plateau in the optical and X-ray flux.  \label{afterglow}}
\end{center}
\end{figure}

Despite the low signal to noise there is apparently little evidence for fading within the first Liverpool Telescope observations which indicate that the afterglow decay was roughly flat, or decaying with $\alpha \sim 0.07 \pm 0.28$ (assuming a power-law decay of the form $F(t) \propto t^{-\alpha}$).  This can be compared to the X-ray decay which shows a plateau during this period (see Fig.\thinspace\ref{afterglow}).

Including our later time WHT observation the optical decay rate becomes $\alpha = 0.86 \pm 0.10$ (However, there is also a significant contribution from the host galaxy at this point which was subtracted but given the relatively small field of view of our WHT/AUX port image, and the difference in the observed filter, a PSF matched subtraction yields large residuals).  This can be compared to the X-ray decay over a similar time frame of $\alpha = 1.73 \pm 0.11$ \citep{GCN6627}.  While the optical points are roughly flat during the x-ray plateau, suggesting that the two regimes were not entirely disconnected, the optical and x-ray behaviors clearly diverge at late times.

\subsection{Photometric redshift} \label{photred}
\begin{figure}[h]
\begin{center}
\includegraphics[width=0.7\textwidth]{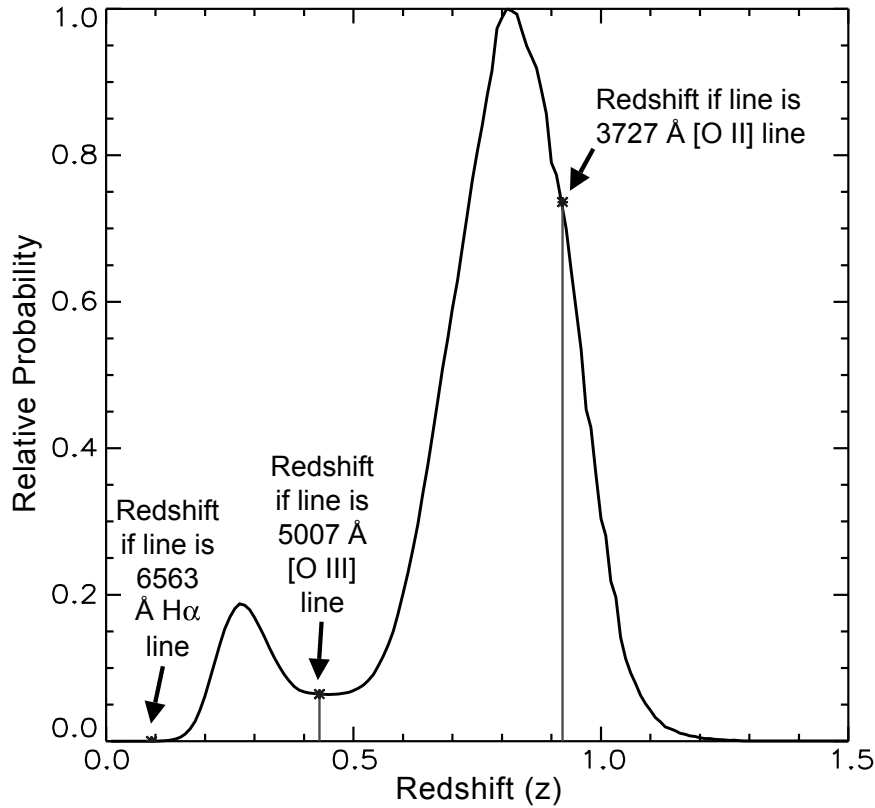}
\caption{Photometric redshift relative probability distribution with the various line possibilities annotated. \label{prob}}
\end{center}
\end{figure}
Our optical and IR photometry also allows us to constrain the 7167 {\AA} line identity via a photometric redshift determination. We calculated a photometric redshift probability distribution for the host galaxy using template-fitting \citep{gwyn, mobasher}. In this method, the observed photometry is matched to synthetic photometry derived using the filter throughputs and a library of galaxy spectral templates red-shifted in the range $0<z<6$. The photometric redshift is derived by minimizing the $\chi^2$~value

\begin{equation}
\chi^2=\sum^7_{n=1}([F^n_{obs}-\alpha F^n_{template}]/\sigma^i)^2,
\end{equation}
where the summation is taken over the seven filters available and $F^n_{obs}$~and $F^n_{template}$~are the observed and synthetic fluxes in band $n$, respectively. Finally, $\alpha$~is a normalization constant and $\sigma^n$~is the flux error in band $n$. The spectral templates used here consists of the E, Sbc, Scd, and Im templates from \cite{Coleman}, together with two starburst templates from \cite{Kinney}. Besides the best-fitting photometric redshift, this method also gives a redshift probability distribution, $P(z)$, where
\begin{equation}
P(z)\propto exp(-\chi^2).
\end{equation}

The best-fitting photometric redshift and the relative probabilities for the three likely line possibilities discussed in Section \ref{candidates} are calculated along with their respective best-fitting host galaxy spectral types.  We find a best-fitting photometric redshift $z=0.83^{+0.12}_{-0.20}$ for an Scd type galaxy.  These results are shown in Table \ref{probs} and Fig.\thinspace\ref{prob}.

\begin{figure}[h]
\begin{center}
\includegraphics[width=1\textwidth]{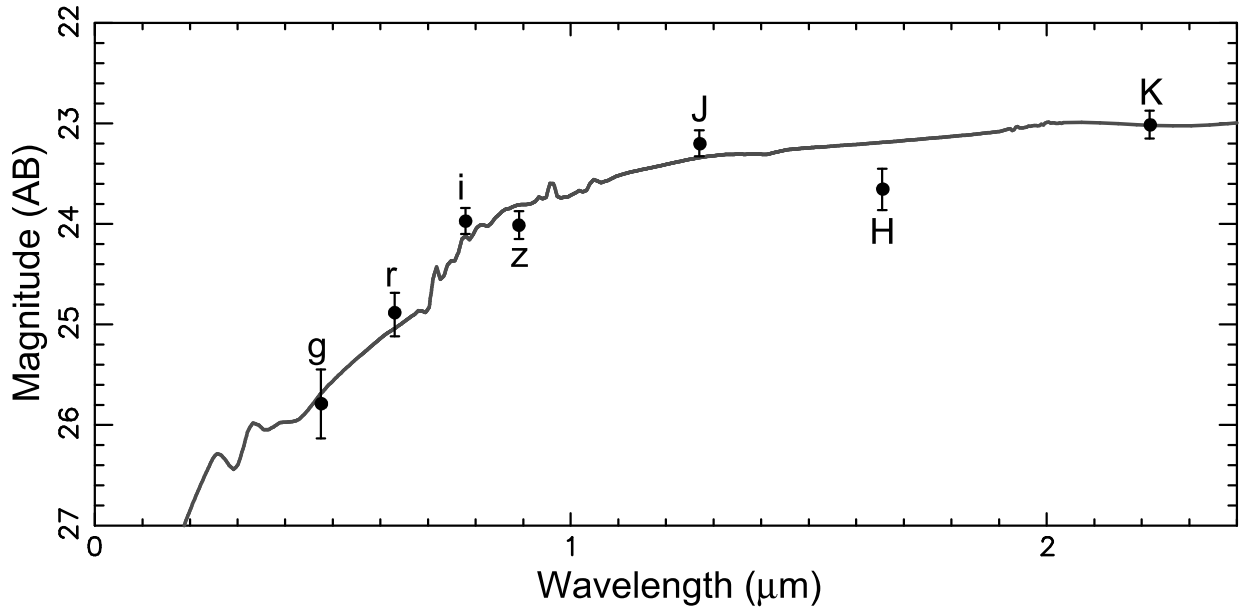}
\caption{The spectral energy distribution template of the z=0.923 fit (solid line) with the measured photometric values over plotted.  (The measured photometric values have been converted to AB magnitudes).  \label{sed}}
\end{center}
\end{figure}

Thus the detected line is by a factor of ten most likely the 3727 {\AA} [O II] line making the photometric redshift consistent with a GRB host galaxy at $z=0.923$.  The best fit spectral energy distribution is with an Scd type galaxy template of magnitude V=-19.7.  Fig.\thinspace\ref{sed} shows the measured photometry plotted over the redshifted template.

\begin{table}[h]
\begin{center}
\begin{tabular}{|c|c|c|c|c|c|c|}

\hline
        Line    &  Redshift  & Relative & Galaxy & Absolute \\
        Matched    &  (z) & Probability & Type & Mag (V) \\
\hline
        Pure Photometric Fit    &  0.83  & 1 & Scd & -19.4\\
        6563 {\AA} H$\alpha$    &  0.09  & 0.004 & Sbc &  -13.8 \\
        5007 {\AA} [O III]    &  0.43  & 0.057 & Scd & -17.8  \\
        3727 {\AA} [O II]      &  0.92  & 0.88 & Scd & -19.7 \\
\hline

\end{tabular}
\caption{Photometric redshift relative probabilities for various line possibilities. \label{probs}}
\end{center}
\end{table}


\subsection{Radial velocity curve and correction}
\begin{figure}[h]
\begin{center}
\includegraphics[width=0.8\textwidth]{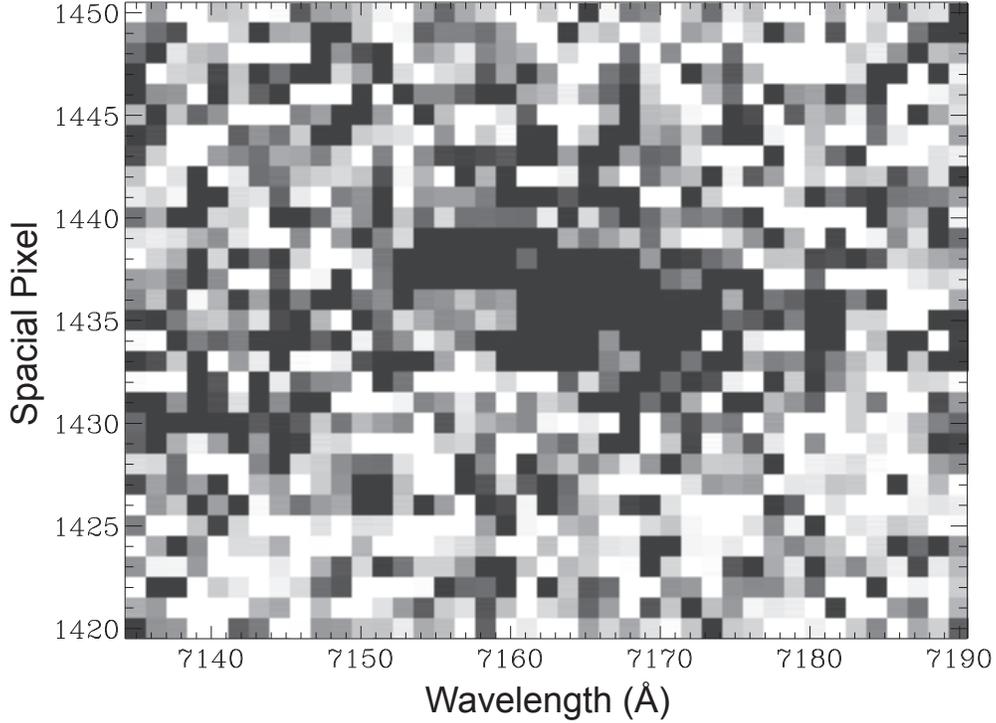}
\caption{Two dimensional spectra image of the 7167 {\AA} line.  Note that the line has a radial velocity skew.  \label{line2d}}
\end{center}
\end{figure}

\begin{figure}[h]
\begin{center}
\includegraphics[width=0.7\textwidth]{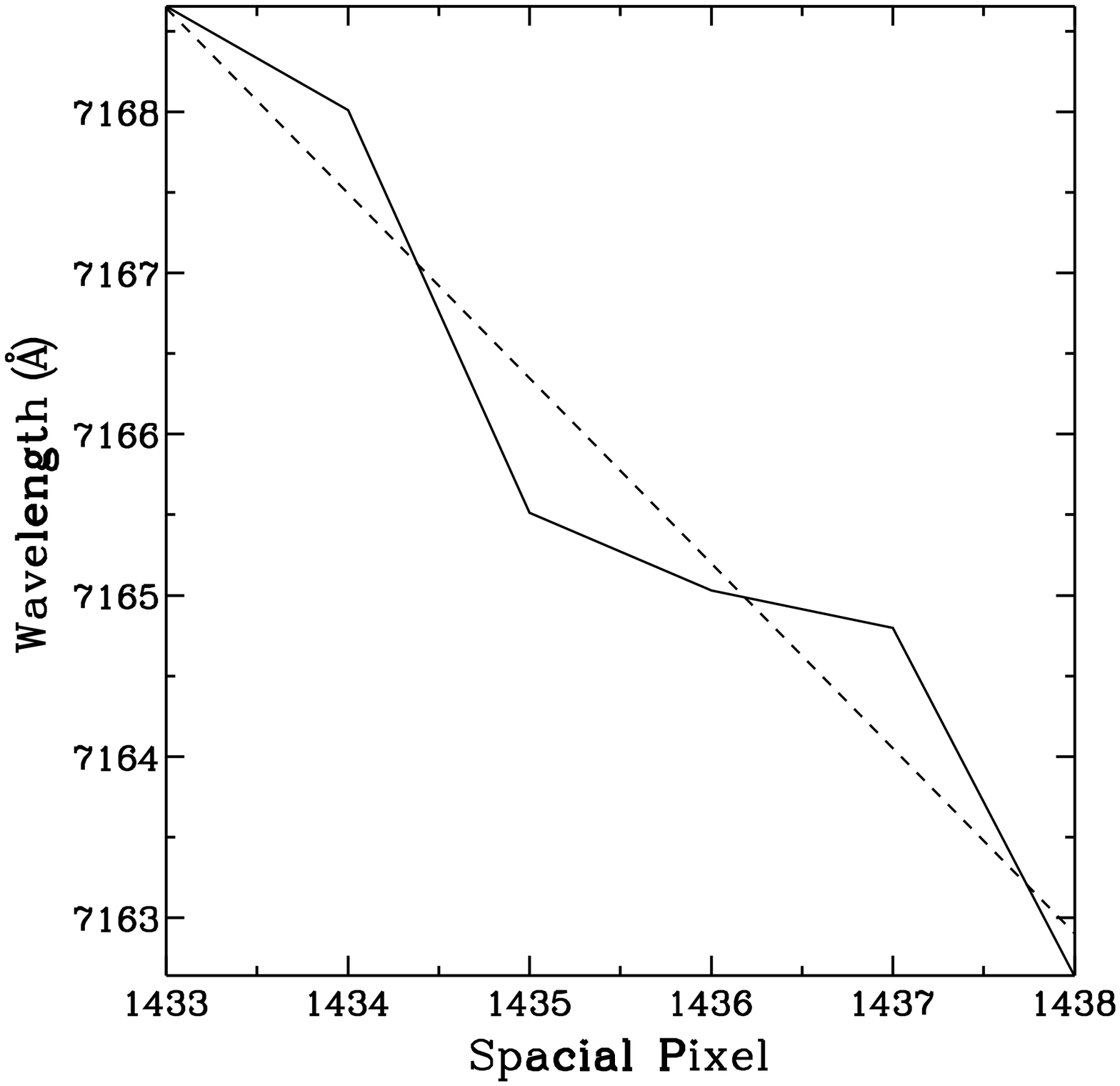}
\caption{Radial velocity plot of the 7167 {\AA} line.  The dotted line shows the fitted radial velocity.  \label{vel_plot}}
\end{center}
\end{figure}
A two dimensional view of the 7167 {\AA} line (see Fig.\thinspace\ref{line2d}) shows an apparent shift in the line wavelength along the spatial direction due to the rotation of the galaxy.  Fitting a radial velocity curve gives an estimated galactic rotation velocity of 110 $\pm$ 20 km s$^{-1}$.  Assuming an M$_{V}$= -19.7 Scd galaxy as suggested by our photometric redshift fitting (see Section \ref{photred}), and a rest-frame color term of B-V = 0.5, applying the redshift dependent B-band Tully-Fisher relation in \cite{bohm} gives a circular velocity of 83 $\pm$ 20 km s$^{-1}$.  (Applying the local Tully-Fisher relation with the same color term gives a circular velocity of 121$\pm$ 15 km s$^{-1}$).  Thus the detected radial velocity curve is reasonably consistent with our photometric redshift fitting.

The radial velocity curve has the effect of spreading out the spectrum during one dimensional extraction effectively decreasing the spectral resolution and correspondingly the achievable signal to noise.  However, on the two dimensional spectrum, the radial velocity curve is separated in spatial direction thus this effect can be removed by shifting each row of the spectrum (in the spectral direction) such that spectral features line up in the spatial direction; effectively removing the rotation curve from the two dimensional spectrum.  Since the radial velocity curve closely approximates a difference of one spectral pixel for each spatial pixel, this shift was employed.  The one dimensional extraction was again performed with the IRAF ``apall" task using identical parameters as previously employed (see Section \ref{apall}).  This counter shifted spectrum is used subsequently and yields an estimated 15\% improvement in signal to noise.

\subsection{Line Flux Measurement}

After smoothing, the continuum was fit with a polynomial and the flux around the line was determined to be $6.39 \pm 0.30 \times 10^{-19}$ erg s$^{-1}$ cm$^{-2}$ {\AA}$^{-1}$. Due to the spectrum being rather noisy Gaussian fitting was thus unusually affected by the initial fitting parameters.  While this fitting yielded values consistent with the line flux determined subsequently (using a method that makes no assumption on the lines shape and width) estimates of the error on the determined flux values were an unacceptably high 20 to 30 \% of the line flux.

In order to obtain a more robust measure of the flux, a curve of growth technique was used in which the spectrum is boxcar summed with increasing widths and the resultant highest value pixel taken as the flux.  With this method the flux will be underestimated until a sufficient width is achieved to encompass all line flux and thereafter higher widths will cause the flux measurements to oscillate, to increasing degree, about the true line flux from the increasing noise added via the addition of non-line pixels into the smoothing.  This approach is equivalent to the "curve-of-growth" technique used in photometry.

An optimal smoothing of 17 pixels yielded a flux with a curve of growth fitting error of $1.664 \times 10^{-18}$ erg s$^{-1}$ cm$^{-2}$.  An additional error of $1.855 \times 10^{-19}$ erg s$^{-1}$ cm$^{-2}$ pixel $^{-1}$, is calculated from the standard deviation of the smoothed spectrum around the line from the continuum.  The flux of the 7167 {\AA} line (in the observer frame) is thus determined to be $2.7 \pm 0.36 \times 10^{-17}$ erg s$^{-1}$ cm$^{-2}$ yielding a S/N ratio of 7.69 and an (galactic rest frame) equivalent width of $-22.3 \pm 2.9$ {\AA} (assuming the line is the 3727 [O II] line).

\subsection{Spectroscopic constraints} \label{candidates}

In addition to photometric redshift fitting is it possible to consider constraints on the identity of the detected line based on the flux limits for non-detected lines.  Given the range of continuous spectral coverage obtained, there are three reasonable candidates for the single observed spectral line at 7165 {\AA}; the 3727 {\AA} [O II] line placing the object at a redshift of z=0.92, the 5007 {\AA} [O III] line placing the object at a redshift of z=0.43, and the 6563 H$\alpha$ line placing the object at a redshift of z=0.09.  The latter possibility is highly unlikely (see Table \ref{probs}) and can be excluded given the color the object alone and thus is not given further consideration.  The 4959 {\AA} [O II] line is also notable but not a candidate due to its flux being quantum mechanically pegged at about a third of the flux of the 5007 {\AA} [O III] line.

Assuming the 7167 {\AA} spectral feature is the 3727 {\AA} [O II] line, the 5007 {\AA} [O III] line would be expected at a detector wavelength of 9628 {\AA} \label{3727->5007}.  While the GMOS sensitivity is quite poor beyond 9250 {\AA}, the expected position was included in the spectral range of the September 13th spectroscopy.  No line was positively detected, however a feature was seen at the expected wavelength with an approximate flux of $5 \times 10^{-17}$ erg s$^{-1}$ cm$^{-2}$ which, while not typical of, is consistent with other noise features in its respective region.  A flux upper limit of $6.7 \times 10^{-18}$ erg s$^{-1}$ cm$^{-2}$ can be placed on the 5007 {\AA} [O III] line.  This yields a lower limit on the 3727 {\AA} [O II] to 5007 {\AA} [O III] line flux ratio of 0.40 $\pm$ 0.19 in the z=0.92 case.  Similarly, assuming the 7167 {\AA} spectral feature is the 5007 {\AA} [O III] line, the 3727 {\AA} [O II] line would be expected at a detector wavelength of 5335 {\AA} (within the spectral range of the July 25th spectroscopy).  No line was detected however a flux limit of $4.0 \times 10^{-17}$ erg s$^{-1}$ cm$^{-2}$ can be placed on the non-detected 3727 {\AA} [O II] line.  This yields an upper limit on the 3727 {\AA} [O II] to 5007 {\AA} [O III] line flux ratio of 1.48 $\pm$ 0.31 in the z=0.43 case.  Since the 3727 {\AA} [O II] to 5007 {\AA} [O III] line flux ratio is degenerately dependent on the extinction, ionization parameter, and metallicity of the galaxy a wide range of ratios have been observed with values ranging from .02 to 25 \citep{kd2002} and without additional information on the galaxy the permissible range can not be constrained (galaxy type is not particularly constraining).  Thus no exclusion can be placed on either possibility based on the 3727 {\AA} [O II] to 5007 {\AA} [O III] line flux ratio.

A more constraining limit on the 5007 {\AA} [O III] line case is that the 4959 {\AA} [O III] line would be expected at 7098 {\AA}.  Again, no line was detected; however a flux limit of $6.8 \times 10^{-18}$ erg s$^{-1}$ cm$^{-2}$ can be placed on the non-detected 4959 {\AA} [O III] line.  Since the 4959 {\AA} [O III] line and the 5007 {\AA} [O III] line are both the result of spontaneous forbidden transitions from the $2s^{2}$ $2p^{2}$ $^{1}D_{2}$ state (to $2s^{2}$ $2p^{2}$ $^{3}P_{1}$ and $2s^{2}$ $2p^{2}$ $^{3}P_{2}$ respectively) they have a quantum mechanically fixed flux ratio of 1:3.01 \citep{Storey} corresponding to a flux intensity ratio of 1:2.98 \citep{Storey}.  This is in reasonable agreement with the observed galactic line flux ratio of 1:2.953$\pm$0.014 \citep{Dimitrijevic}.  Thus the expected flux of the 4959 {\AA} [O III] line is $9.06 \pm 1.21 \times 10^{-18}$ erg s$^{-1}$ cm$^{-2}$ and quite well constrained.  This expected flux marginally exceeds the observed flux limit.

For the 3727 {\AA} [O II] line case, near infrared spectroscopy was performed in an effort to detect the H$\alpha$ line shifted out to 1.26 $\mu$m.   No line was detected; however a flux limit of $ 4.4 \times 10^{-17}$ erg s$^{-1}$ cm$^{-2}$ can be placed on the non-detected line.  Due to the expected observer frame H$\alpha$ wavelength being close to the J band central wavelength we can make a crude estimate on the continuum flux based on our J band IR imaging.  This gives an estimated continuum flux of $3.7 \times 10^{-19}$ erg s$^{-1}$ cm$^{-2}$ {\AA}$^{-1}$ and a corresponding approximate upper limit on the H$\alpha$ (galactic rest frame) equivalent width of 62 {\AA}.  This yields a lower limit on the 3727 {\AA} [O II] to H$\alpha$ ratio in flux of 0.45 and an equivalent width of 0.36 which, given that a majority of galaxies are within these constraints, makes the non-detection of the H$\alpha$ line perfectly reasonable.

Thus, from the spectroscopic data alone the 3727 {\AA} [O II] line is marginally favored based on the exclusion of the 5007 {\AA} [O III] case from the non-detection of the 4959 {\AA} [O III] line at the quantum mechanically required flux ratio.  However in concert with photometric redshift fitting which strongly favors 3727 {\AA} [O II] as the line, this can be reasonably assumed to be the correct identification and yields a redshift of z=0.923 for the host galaxy.

\section{Conclusions}

Initial \emph{Swift} observations put GRB 070714B securely in the short burst (SGRB) category.  Observations with the Liverpool Telescope detected an optical afterglow with an initial plateau for the first 5 to 25 minutes, then subsequent decay steepening to $\alpha = 0.86 \pm 0.10$ for the remaining first 24 hours post burst.  This is consistent with the X-ray light-curve which also shows an initial plateau followed by concurrent decay of $\alpha = 1.73 \pm 0.11$ \citep{GCN6627}, suggesting that the X-ray and optical regimes were not entirely disconnected.

We also detect a host galaxy with an angular extent that includes the location of the optical transient (0.4" from the host galaxy center).  Our Gemini Nod \& Shuffle spectroscopic observations of the host show a single emission line at 7167 {\AA} which, based on a photometric redshift from our \textit{grizJHK} multi-band optical and infrared photometry implies this can only be the 3727 {\AA} [O II] line.  This places the host at a redshift of z=0.923, the highest spectroscopically confirmed redshift for a short burst.

Photometric fitting shows the host to posses a type Scd stellar population and its luminosity, color, and radial velocity are all consistent with an Scd and are in fact quite similar to M33.  The host's diameter, about 7 to 8 kpc, is notably only about half of M33's; however given the redshift of the object (z = 0.92) and that grand design spirals only emerge at around z of 1 this difference in angular size does not seem that problematic.  Planned HST imaging should allow us to tell whether the similarity between colors and velocity extends to morphology.

Subsequent to our discovery of the redshift of SGRB 070714B the host galaxy of SGRB 070429B was observed with Keck to have a spectral line at 7091 {\AA}.  This, the observers suggest, was the 3727 {\AA} [O II] line, placing it at a redshift of z=0.904 \citep{GCN7140, Berger}.  However, in the absence of either a photometric redshift or other spectral features, it is hard to estimate the likelihood that this line identification is correct.  Prior to these discoveries the highest spectroscopically confirmed short burst redshift was SGRB 051221 at z=0.546 \citep{Soderberg2006}.

A number of observers have obtained the redshifts of bright galaxies found within XRT error circles, and have suggested that these may be used as the redshifts of the bursts \citep{Berger2007}.  However, \cite{CobbBailyn} have shown that if an $I  < 21.5$ magnitude galaxy is detected in an XRT error circle there is a 50\% chance that it is due to random alignment of an unrelated galaxy.  In contrast, they expect a contamination rate of only 1\% based on the detection of an optical transient within the angular extent of  galaxy.  Indeed, even a radius of 5" (consistent with many claimed associations and typical of the pre UVOT enhanced XRT position error circles) would have a greater than $60\%$ chance of containing a random galaxy as bright as the host of GRB 070714B based on HDF galaxy counts \citep{HDF}.  Thus, only in cases such as GRBs 070714B and 070429B where the burst has an optical afterglow within the angular extent of a galaxy can the host association be made with very little chance of confusion \citep{LevanHB}.  We are fortunate in the case of 070714B to be able to further strengthen our single-line redshift measurement by a highly-constraining photometric redshift.

While 070714B is now the short burst with the highest spectroscopic redshift, and alone extends examples of short burst formation throughout the recent half of the universe, there are other short bursts that may well be more distant.  \cite{Berger2007} has pointed out that there are no bright galaxies in the XRT error circles of a number of bursts, suggesting that at least some of these are at redshifts $z > 1$  and indeed there are cases like GRB 060121 \citep{GCN4841} where deep observations have turned up faint hosts not initially detected.  Furthermore, photometric observations of the afterglow and host of GRB 060121 suggest it was probably above $z \sim 1$, and may have been at $z \sim 4.5$ \citep{Donaghy, dUP, Levan060121}.   Recent additional HST and SPITZER observations of the host of SGRB 060121 may produce a more reliable photometric redshift estimate (Levan et al. in prep).  Additionally, the possibility that the highest redshift GRB (GRB 080913 at z = 6.7) {\it might} be categorized as a short burst would push short burst formation even into the universe's first billion years \citep{Belczynski080913, Greiner080913, Zhang080913}.

From our redshift and the observed burst fluence (of $7.2 \times 10^{-7}$ erg cm$^{-2}$ from GCN 6623 \citealt{GCN6623}) we calculate an isotropic energy release (Eiso) for SGRB 070714B of $1.61 \pm 0.20 \times 10^{51}$ erg (in the \emph{Swift} bandpass pass).  \cite{Kanninpress} (using the method outlined in \citealt{bolometric}) estimates a bolometric isotropic energy release (Eiso) of  $9.8^{+4.0}_{-2.0} \times 10^{51}$ erg which is in reasonable agreement with a  15-2000 keV band value of $8.3^{+2.9}_{-1.3} \times 10^{51}$ erg calculated from a joint \emph{Swift}\thinspace+\thinspace\emph{Suzaku-WAM} spectral analysis fluence value of $3.7^{+1.3}_{-0.6} \times 10^{-6}$ erg cm$^{-2}$ \citep{GCN6638}.  This places SGRB 070714B about an order of magnitude more luminous than the median short burst yet still about an order of magnitude below the luminosity of the median long burst (using the data compiled in \citealt{Nysewander}).  SGRB 070714B has almost twice the energy release of the previous highest redshift short burst SGRB 051221 and makes it the most luminous short burst (from the sample of those with spectroscopic redshifts) yet seen \citep{Nysewander, Kanninpress}, though still more than an order of magnitude lower than SGRB 060121 even at the lower of the two photometric redshift solutions quoted in \cite{dUP}.

\begin{figure}[h]
\begin{center}
\includegraphics[width=01\textwidth]{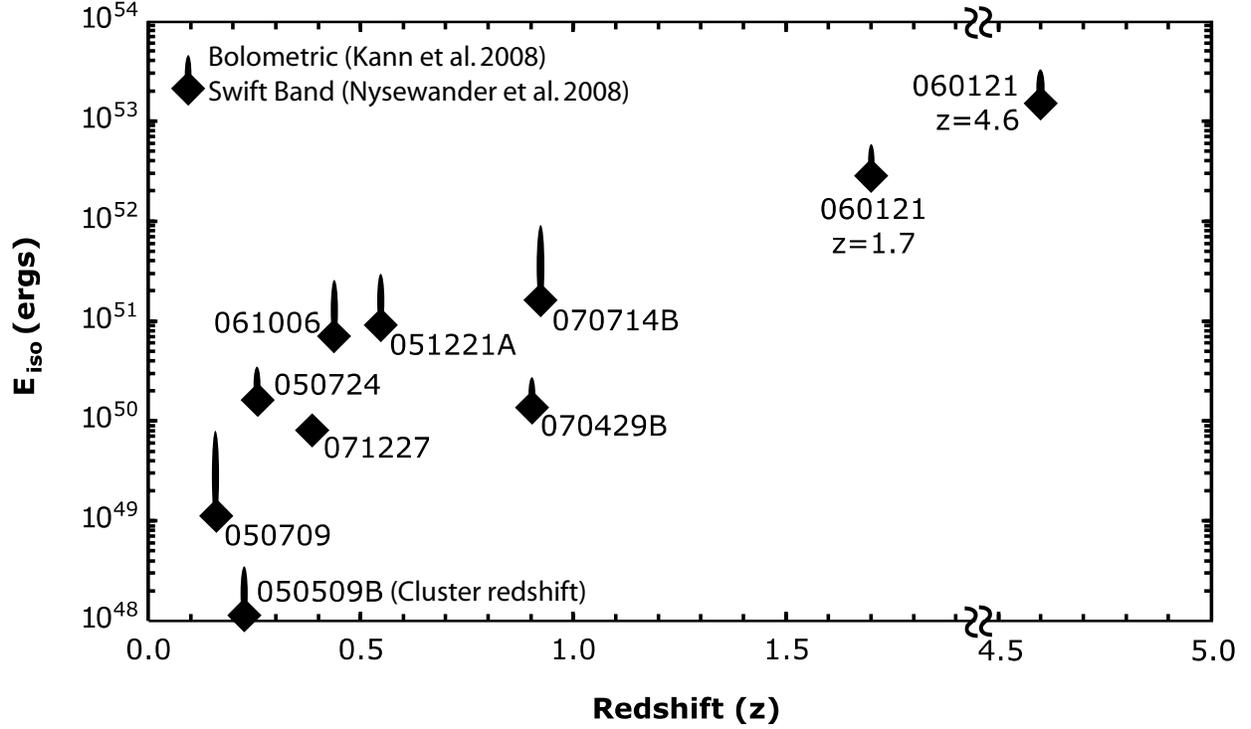}
\caption{Plot of E$_{iso}$ vs. redshift for short bursts.  In order to insure accuracy the sample is limited to those bursts with redshifts derived spectroscopically from galaxies in which an optical afterglow was located and GRB 050509B (whose extreme proximity to a rich cluster makes an association highly probable). The two photometric redshift possibilities for SGRB 060121 are both plotted and specific SGRBs are annotated.  Values in the \emph{Swift} bandpass, from \citep{Nysewander}, are shown (diamonds) with estimated bolometric corrections from \citep{Kanninpress} applied (lines aboive diamonds) where available (all cases except GRB 071227). \label{eisoz}}
\end{center}
\end{figure}

There appears to be significant bias at the highest redshifts toward preferentially detecting the most luminous short bursts (see Fig.\thinspace\ref{eisoz}).  In particular, with the noted exception of SGRB 070429B,  the highest redshift short bursts consistently have a burst luminosity higher than any lower redshift burst.  Whether this trend of finding ever more luminous bursts as we probe ever higher redshifts is partially due to evolution in burst luminosity with redshift or solely the result of detector sensitivity limits and other selection effects remains to be determined.  Regardless, considerable care must be exercised when studying the properties of high redshift short bursts and especially when comparing them with those of the local population.

SGRB 070714B, and also SGRB 070429B, firmly moves back the time at which we know short bursts were being formed into the first half of the age of the universe (and further dispels the notion that short bursts require an old progenitor population).  With further host identifications, it may become possible to observe evolution in the SGRB host galaxy types and possibly set observational constraints on the formation time required by the progenitor system.  Indeed in the case of Type Ia supernovae, a lack of of type Ia's observed at z $>$ 1.4, and the suggested dearth thereof, has put considerable constraints on their formation time and models \citep{Strolger, Dahlen}.  The ever increasing redshift detections, and likely bias towards detection of only the more luminous short bursts at the highest redshifts, suggests that short bursts may well have occurred throughout nearly all the history of the universe.

\acknowledgments
This paper is based on observations made with the Gemini North and South, KECK II, Liverpool, and William Herschel Telescopes.

We wish to thank Kathy Roth and other members of the Gemini staff for their help in scheduling and implementing these observations.  The Gemini Observatory is operated by the Association of Universities for Research in Astronomy, Inc., under a cooperative agreement with the NSF on behalf of the Gemini partnership: the National Science Foundation (United States), the Science and Technology Facilities Council (United Kingdom), the National Research Council (Canada), CONICYT (Chile), the Australian Research Council (Australia), MinistŽrio da Cincia e Tecnologia (Brazil) and SECYT (Argentina).  Data was collected under programs  GN-2007A-Q-28, GN-2007B-Q-31, and GS-2007B-Q-12.

The W.M. Keck Observatory is operated as a scientific partnership among the California Institute of Technology, the University of California and the National Aeronautics and Space Administration.  The  Observatory was made possible by the generous financial support of the  W.M. Keck Foundation.  The authors wish to recognize and acknowledge the very significant cultural role and reverence that the summit of Mauna Kea has always had within the indigenous Hawaiian community.  We are most fortunate to have the opportunity to conduct observations from this mountain.

The Liverpool Telescope is operated on the island of La Palma by Liverpool John Moores University in the Spanish Observatorio del Roque de los Muchachos of the Instituto de Astrofisica de Canarias with financial support from the UK Science and Technology Facilities Council.

The William Herschel Telescope is operated on the island of La Palma by the Isaac Newton Group in the Spanish Observatorio del Roque de los Muchachos of the Instituto de Astrofisica de Canarias.  We thank the overridden observes David Bonfield and Alejo Sansigre for performing our ToO observations.

{\it Facilities:} \facility{Gemini:Gillett (GMOS, NIRI)}, \facility{Keck:II (NIRSPEC)}, \facility{Gemini:South (GMOS)},  \facility{Liverpool:2m}, \facility{ING:Herschel}

\bibliographystyle{apj}
\bibliography{\jobname}

\end{document}